\documentclass[aps,
prx,
twocolumn,
superscriptaddress,
nofootinbib,
floatfix,
longbibliography]{revtex4-2}

\usepackage{amsmath,amssymb,bm}
\usepackage{graphicx}
\usepackage{booktabs}
\usepackage{array,microtype}
\usepackage[colorlinks=true,citecolor=blue,linkcolor=blue,urlcolor=blue]{hyperref}
\usepackage{tikz}

\newcommand{\kcav}{\kappa_{\mathrm{cav}}}
\newcommand{\kph}{\kappa_{\mathrm{ph}}}
\newcommand{\RuCl}{\ensuremath{\alpha\text{-RuCl}_{3}}}
\newcommand{\Pzero}{\mathcal P_0}
\newcommand{\Qzero}{\mathcal Q_0}
\newcommand{\Dop}{\Delta_{\mathrm{op}}}
\newcommand{\Othree}{O_{3}}
\newcommand{\cminus}{c_{-}}

\tikzset{
  panel label/.style={
    anchor=north west,
    font=\footnotesize,
    fill=white,
    inner sep=1.5pt,
    rounded corners=1pt
  }
}

\newcommand{\mytitle}{Chiral Cavity Vacua as Local Generators and Probes of Non-Abelian Kitaev Topology}

\begin{document}

\title{\mytitle}
\author{Yu Zhang}
\email{zhy@lanl.gov}
\affiliation{Theoretical Division, Los Alamos National Laboratory, Los Alamos, New Mexico 87545, USA}
\date{\today}

\begin{abstract}
We show how a gyrotropic cavity vacuum generates, probes, and spatially programs non-Abelian topology in the Kitaev honeycomb model. Gauge-consistent virtual-photon elimination produces a universal helicity-odd interaction $\mathrm{i}[P_x,P_y]$ alongside same-order even dressing. For the leading bond polarization, this commutator is exactly the Kitaev three-spin mass. Mode normalization imposes a thermodynamic constraint: one extended normalized mode gives a local mass that vanishes with sample area, whereas a localized mode induces a finite topological island and a finite mode density enables a bulk mass. The exact Majorana solution yields helicity-reversed Chern gaps and edge propagation, while integrated circular dichroism exactly measures the mass generator, $\langle\mathrm{i}[P_x,P_y]\rangle=-\sqrt3p^2\langle O_3\rangle$. Real-space calculations reveal a non-Abelian island and a helicity wall with two co-propagating Majorana modes. Uniform helicity reversal is a Dirac-mass quench with $n_{\rm ex}\propto\tau_Q^{-1}$, whereas moving a wall localizes the gap closing. The equilibrium vacuum kernel contains sum-frequency rather than resonant denominators. These results establish distinct criteria for local and bulk cavity-induced topology.
\end{abstract}

\maketitle

Quantum spin liquids support fractionalized excitations and emergent gauge structure without conventional magnetic order~\cite{Anderson1973,Balents2010,SavaryBalents2017,Broholm2020,KnolleMoessner2019}. The Kitaev honeycomb model is an unusually sharp platform because spin-$1/2$ moments fractionalize into Majorana fermions coupled to a static $Z_2$ gauge field, and a time-reversal-breaking three-spin interaction gaps the Dirac cones into an Ising topological phase with Majorana Chern number $|\nu_M|=1$ and chiral central charge $|\cminus|=1/2$~\cite{Kitaev2006,ReadGreen2000,Nayak2008,Hermanns2018}. Its dynamical response provides characteristic signatures of the emergent Majorana fermions and gauge fluxes~\cite{Knolle2014}. This structure motivates the search for Kitaev materials, including \RuCl, but quantitative interpretation is complicated by non-Kitaev exchange, magnetic order, phonons, and sample dependence~\cite{JackeliKhaliullin2009,Rau2014,Winter2017,TrebstHickey2022,Plumb2014,Banerjee2016,Kasahara2018,Czajka2021,Bruin2022}.

Cavity quantum materials offer a complementary control knob: electromagnetic vacuum fluctuations can alter equilibrium many-body Hamiltonians without a continuously external drive~\cite{Ruggenthaler2018,Schlawin2022,FornDiaz2019,Sentef2018,Kirton2019,seshadri2026cavitymott, aklilu2026arxiv, taylor2026bloch, Weight:2023aa, Bauman:2025aa}. Cavity magnonics further demonstrates the broader opportunity to couple confined electromagnetic fields to collective magnetic excitations~\cite{Rameshti2022}. Cavities have been proposed to generate frustrated magnetic interactions and spin liquids~\cite{Chiocchetta2021,Fadler2024}, to reshape the magnetic phase diagram of {\RuCl}~\cite{Bostrom2023}, and to stabilize an Abelian kagome chiral spin liquid~\cite{Wei2025}. Chiral terahertz cavities with polarization-selective modes and broken time-reversal symmetry have progressed from realistic design to experiment~\cite{Tay2025,Kulkarni2026}.

The central operator question is whether a stationary chiral vacuum can generate the \emph{specific} mass of the non-Abelian Kitaev phase. A second, equally important question is spatial: does one normalized cavity mode produce an intensive thermodynamic mass, or only a finite-size/local effect? Third: can the same light--matter vertex provide a quantitative probe of the induced order? The present work answers all three questions in an exactly controlled setting.

Our starting point is a vacuum Schrieffer--Wolff identity~\cite{SchriefferWolff1966,BravyiDiVincenzoLoss2011}. A circular mode couples to two projected polarization components, and virtual-photon elimination converts their noncommutativity into a helicity-odd Hamiltonian. We then keep the mode profile rather than replacing it by a uniform coupling. This exposes a participation theorem that distinguishes a single extended mode, a localized nanocavity, and a finite-density multimode environment. For the ideal bond polarization, the resulting local operator is the Kitaev mass exactly. The exact solvable model then supplies the band topology, a Circular Dichroism (CD) sum rule, real-space topological islands and helicity domain walls, switching dynamics, and the full matter-continuum vacuum kernel.

We start from a Kitaev spin Hamiltonian $H_s$, obtained after charge excitations have been downfolded, and couple its projected electric polarization to a localized circular cavity mode:
\begin{align}
H= &H_s+\hbar\omega_c a^\dagger a
+g\left(aQ_\eta^\dagger+a^\dagger Q_\eta\right)+H_{\rm DSE},\nonumber\\
Q_\eta= &\frac{P_x+\mathrm{i}\eta P_y}{\sqrt2},
\quad \eta=\pm1.
\label{eq:lightmatter}
\end{align}
The dipole self-energy (DSE) $H_{\rm DSE}$ must be retained for gauge-consistent quantitative calculations~\cite{DeBernardis2018,Schaefer2020,Dmytruk2021,Ashida2021}. Projecting onto the photon vacuum to second order gives the exact operator resolvent $\delta H_\eta^{(2)}=-g^2 Q_\eta^\dagger\frac{1}{\hbar\omega_c+H_s-E}Q_\eta$. For a mode dispersive with respect to the spin transitions sampled by $P_a$,
\begin{align}
\delta H_\eta^{(2)}= &-\lambda(P_x^2+P_y^2)
-\eta\lambda\mathrm{i}[P_x,P_y]+O(\omega_c^{-2}),
\label{eq:swidentity}
\end{align}
where $\lambda=\frac{g^2}{2\hbar\omega_c}$ characterizes the coupling strength.
The first term is time-reversal even. The second is Hermitian, time-reversal odd (denoted as $H_{\rm vac}^{\rm odd}$), and changes sign with photonic helicity $\eta$. The DSE modifies the helicity-even sector at this order but cannot remove an antisymmetric helicity-odd commutator. A direct three-spin--photon diagonalization verifies Eq.~\eqref{eq:swidentity} to an operator-norm error $4.5\times10^{-16}$ and the residual low-energy spectral error scales as $(g/\omega_c)^4$ (Supplemental Material).

Let the projected polarization be resolved into local bond contributions, $P_a[u]=\sum_b u_b P_{a,b},\quad \sum_b |u_b|^2=1$, where the normalization follows from the electromagnetic energy normalization of one mode. The odd interaction is $\mathrm{i}[P_x[u],P_y[u]]=\sum_{bb'}u_b^*u_{b'} \mathrm{i}[P_{x,b},P_{y,b'}]$.
Local bond operators commute unless they overlap. Consequently, the coefficient of a three-spin term centered at site $j$ is a finite sum of products such as $u_{b(j)}^*u_{b'(j)}$.

\begin{figure}[tbp]
\centering
  \begin{tikzpicture}
    \node[inner sep=0] (imgA) {\includegraphics[width=.555\linewidth]{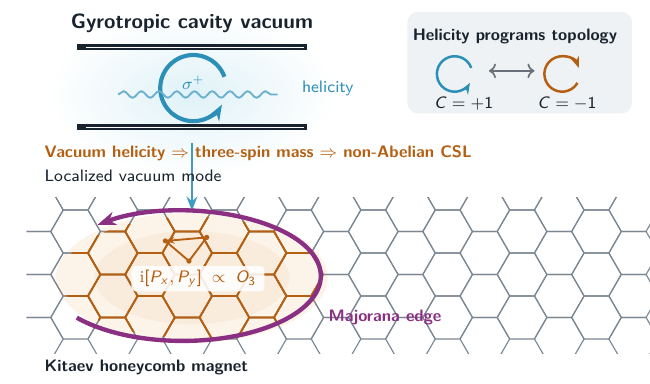}};
    \node[panel label, xshift=2pt, yshift=-2pt] at (imgA.north west) {a)};
  \end{tikzpicture}
  \begin{tikzpicture}
    \node[inner sep=0] (imgA) {\includegraphics[width=.415\linewidth]{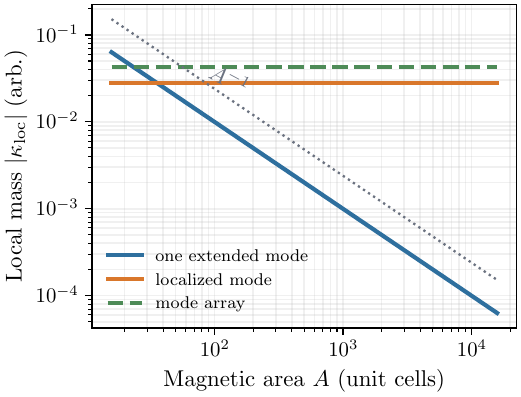}};
    \node[panel label, xshift=-1pt, yshift=0pt] at (imgA.north west) {b)};
  \end{tikzpicture}
\caption{\label{fig:participation}Cavity-written local Ising topology and normalized-mode participation. (a) A localized gyrotropic vacuum generates the helicity-odd commutator $\mathrm{i}[P_x,P_y]$, which becomes the Kitaev three-spin mass $O_3$ and opens a non-Abelian island inside the illuminated region. Reversing the cavity helicity reverses the Majorana Chern number and chiral boundary propagation. (b) The local mass from one extended normalized mode vanishes as $A^{-1}$, while a localized mode retains a finite mass in its footprint and an array at fixed areal density produces an intensive bulk scale. The vertical offsets are illustrative. The scaling with area is the physical content.}
\end{figure}

For a laterally extended mode over $N_b\propto A$ bonds, $|u_b|\sim N_b^{-1/2}$ and therefore $\kappa_j^{(1 \mathrm{mode})}\sim A^{-1}$,
The spatially integrated odd shift remains $O(1)$, but the local mass and bulk gap vanish in the thermodynamic limit. A mode localized over a fixed participation volume $N_{\rm eff}$ instead gives $\kappa_j\sim N_{\rm eff}^{-1}$ within that region and writes a finite topological droplet. An intensive bulk mass requires a number of localized modes proportional to area, or a multimode environment whose local chiral spectral density stays finite as $A\rightarrow\infty$.
For several modes the local kernel is schematically
\begin{equation}
\kappa_j\propto
\sum_\lambda \frac{g_\lambda^2\eta_\lambda}{\hbar\omega_\lambda}
F_j[u_\lambda],
\label{eq:multimode_local}
\end{equation}
with $F_j$ determined by overlapping bonds. Eq.~\eqref{eq:multimode_local}, rather than a collective Rabi splitting alone, is the quantity that decides whether the cavity creates a bulk phase or a local device.
Fig.~\ref{fig:participation} combines the most direct single-cavity implementation with the thermodynamic scaling test. A localized gyrotropic vacuum writes a finite three-spin mass within its participation region, producing a non-Abelian island whose Majorana edge propagation reverses with helicity. In contrast, a single extended normalized mode loses its local mass as $A^{-1}$ according to the scaling law, whereas a localized mode or finite-density mode array retains a finite local scale. An intensive bulk mass specifically requires the latter finite local density of chiral modes, as in Eq.~\eqref{eq:multimode_local}.

Virtual charge transfer in a spin-orbit Mott insulator produces low-energy polarization operators built from spin bilinears~\cite{Resta1994,SouzaWilkensMartin2000,Katsura2005,Bulaevskii2008}. The ideal honeycomb component is
\begin{equation}
\bm P^{\rm spin}=p\sum_{\langle ij\rangle_\gamma}
-\bm\delta_\gamma \sigma_i^\gamma\sigma_j^\gamma+\delta\bm P,
\label{eq:bondpolarization}
\end{equation}
where $\sigma^\gamma$ are Pauli operators in the Kramers doublet, $\bm\delta_\gamma$ are oriented bond vectors, and $p$ is a material-dependent projected dipole. Defining $B_\gamma=\sum_{\langle ij\rangle_\gamma}\sigma_i^\gamma\sigma_j^\gamma$, direct commutator algebra gives $\mathrm{i}[P_x,P_y]=-\sqrt3p^2\Othree$, where
\begin{align}
\Othree&=-\frac{\mathrm{i}}{2}\left([B_x,B_y]+[B_y,B_z]+[B_z,B_x]\right).
\end{align}
Only adjacent bonds contribute, and each term in $\Othree$ is an oriented product $\sigma_i^\alpha\sigma_j^\gamma\sigma_k^\beta$. Therefore
\begin{equation}
H_{\rm vac}^{\rm odd}=\kcav\Othree,
\quad
\kcav=\eta\frac{\sqrt3g^2p^2}{2\hbar\omega_c}
\label{eq:kcav}
\end{equation}
for the stated convention. Symmetry can force $p=0$, an inversion-breaking substrate, gate, interface, asymmetric stack, or an intrinsically odd-parity charge-transfer channel is then required. The full tensor contained in $\delta\bm P$ also generates additional even and odd operators and must be retained in a material calculation.

In the flux-free sector, we write the Kitaev exchange amplitudes as $J_\gamma$ and use $J$ as their balanced energy scale,
\begin{equation}
H_{\rm Kt}=-\sum_{\langle ij\rangle_\gamma}J_\gamma\sigma_i^\gamma\sigma_j^\gamma
+\kcav\Othree
\end{equation}
maps to a two-band Majorana problem
\begin{equation}
\mathcal H(\bm k)=
\begin{pmatrix}d_z(\bm k)&f^*(\bm k)\\f(\bm k)&-d_z(\bm k)\end{pmatrix},
\label{eq:bloch}
\end{equation}
where $f=J_z+J_xe^{\mathrm{i} k_1}+J_ye^{\mathrm{i} k_2}$, $d_z=2\kcav[\sin k_1-\sin k_2+\sin(k_2-k_1)]$.
The Kitaev $B$ phase is gapless when the three $|J_\gamma|$ satisfy the triangle inequalities. Any nonzero $\kcav$ then gives opposite masses at the two valleys and $|\nu_M|=1$. For $J_x=J_y=J$ and $r=J_z/J<2$, the weak-mass gap is
\begin{equation}
\Delta_M=8|\kcav|\left(1+\frac r2\right)\sqrt{1-\frac{r^2}{4}},
\label{eq:anigap}
\end{equation}
which is maximal at the isotropic point, $\Delta_M=6\sqrt3|\kcav|$.
At $r=2$ the cones merge at a zero of the chiral form factor. The preexisting Abelian $A$ phase at $r>2$ is topologically trivial for an infinitesimal mass.

\begin{figure}[t]
\includegraphics[width=\linewidth]{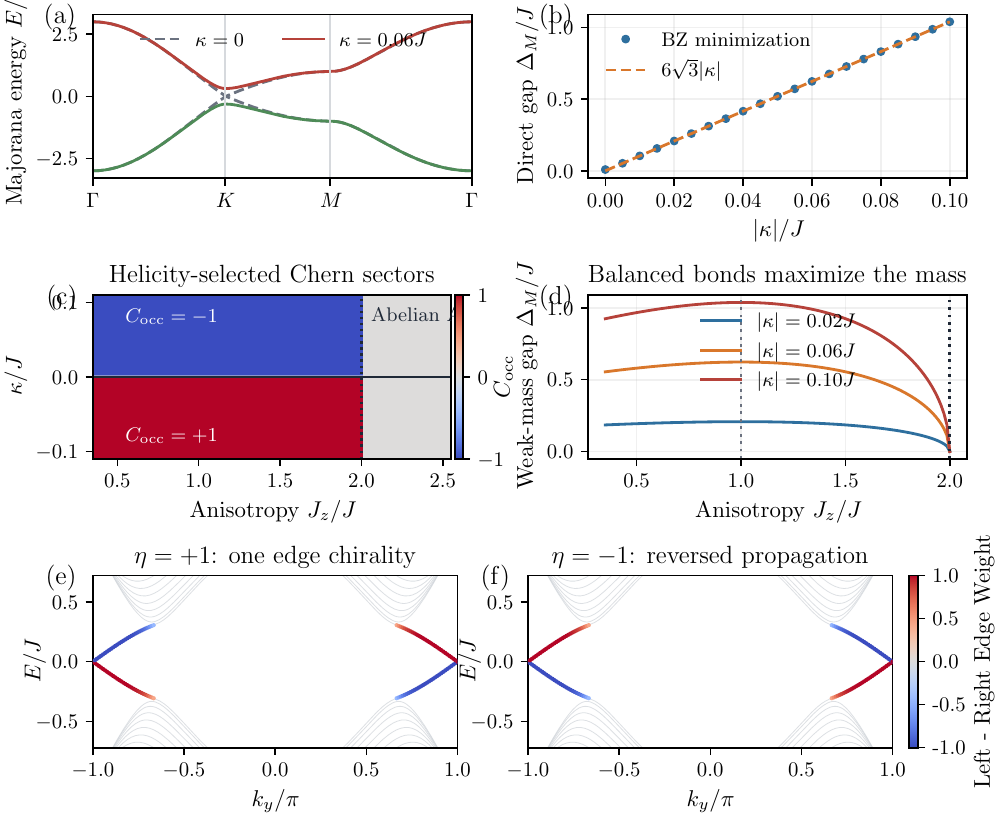}
\caption{\label{fig:topology}Exact non-Abelian topology. (a) The cavity term gaps the two Majorana cones. (b) Brillouin-zone minimization agrees with $\Delta_M=6\sqrt3|\kappa|$ at the isotropic point. (c) Opposite signs of the mass select opposite occupied-band Chern numbers in the Kitaev $B$ region. The strongly anisotropic $A$ region is trivial for an infinitesimal mass. (d) The weak-mass coefficient is maximal for balanced Kitaev bonds and vanishes at cone merger. (e),(f) Ribbon spectra show reversal of the edge propagation under helicity reversal while all bulk eigenvalues remain unchanged.}
\end{figure}

Fig.~\ref{fig:topology} verifies the gap, lattice Chern number, anisotropic window, and bulk--edge correspondence~\cite{Hatsugai1993,Fukui2005}. For the reciprocal orientation adopted here, $C_{\rm occ}=-\operatorname{sgn}(J_xJ_yJ_z\kcav)$. The physical statement independent of orientation is
$\nu_M(\eta)=-\nu_M(-\eta), \cminus=\nu_M/2$.
Opposite helicities have identical bulk energies but reversed neutral edge-energy flow.
The panel-by-panel tests of the mass normalization, anisotropic topological window, and helicity-reversible edge branch are discussed in the Supplemental Material.

The same projected polarization that creates the vacuum mass also provides a direct optical observable. Define the positive-frequency circular absorption spectrum at zero temperature, $I_\eta(\omega)=\sum_{n\ne0} |\langle n|Q_\eta|0\rangle|^2\delta[\omega-(E_n-E_0)/\hbar]$.
Completeness and $Q_\eta=(P_x+\mathrm{i}\eta P_y)/\sqrt2$ give the exact unweighted sum rule
\begin{equation}
\Delta A \equiv \int_0^\infty\mathrm{d}\omega [I_+(\omega)-I_-(\omega)]
=\langle\mathrm{i}[P_x,P_y]\rangle.
\label{eq:sumrule}
\end{equation}
The elastic matrix elements cancel because $P_x$ and $P_y$ are Hermitian. For Eq.~\eqref{eq:bondpolarization},
$\Delta A =-\sqrt3p^2\langle\Othree\rangle =-\sqrt3p^2\frac{\partial E_0}{\partial\kcav}$.
Thus the integrated CD measures the operator conjugate to the cavity-induced topological mass. This is a generation--detection reciprocity: the antisymmetric polarization algebra both generates and diagnoses the chiral interaction.

For the ideal Kitaev polarization, the bond-energy vertex preserves the flux sector and creates a two-Majorana excitation. The exact absorption threshold is therefore the itinerant Majorana gap. At $\kcav=0.06J$, our momentum integration gives a threshold $0.62384J$ and $\Delta A=0.40348947 =-\sqrt3\langle\Othree\rangle$
per unit cell for $p=1$, with an absolute numerical difference $5.6\times10^{-17}$.

\begin{figure}[tbp]
\includegraphics[width=\linewidth]{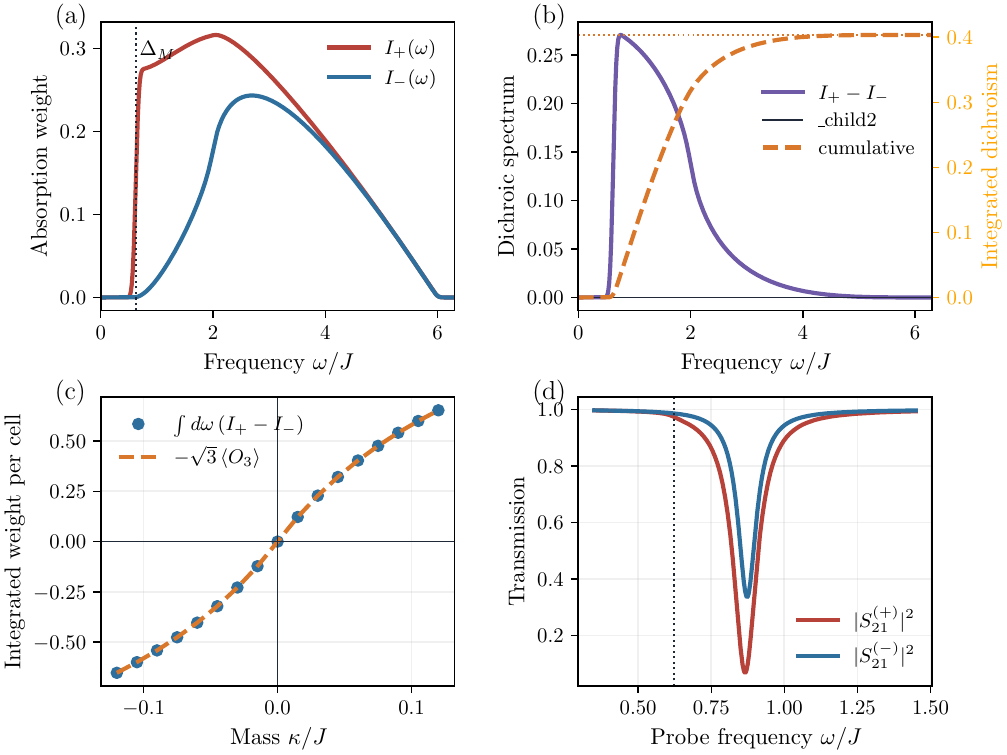}
\caption{\label{fig:spectroscopy}Exact generation--detection reciprocity. (a) Helicity-resolved two-Majorana absorption has a threshold at the cavity-induced gap. (b) The cumulative CD spectrum saturates at $\langle\mathrm{i}[P_x,P_y]\rangle$. (c) Over a mass scan, the integrated dichroism agrees with $-\sqrt3\langle O_3\rangle$ to numerical precision. (d) A representative weak-probe cavity transmission resolves the helicity-dependent matter susceptibility. $\kappa=0.06 J$.
}
\end{figure}

For a weak probe mode, the same spectral samples define the retarded susceptibility. A standard input--output denominator can be written schematically as~\cite{GardinerCollett1985}
\begin{equation}
D_\eta(\omega)=\mathrm{i}(\omega_c-\omega)+\frac{\kph}{2}
+\mathrm{i} g_{\rm pr}^2\chi_\eta^R(\omega),
\quad
S_{21}^{(\eta)}=1-\frac{\kappa_{\rm ext}}{D_\eta}.
\label{eq:inputoutput}
\end{equation}
Here $D_\eta^{-1}$ is the dressed cavity response, $S_{21}^{(\eta)}$ is the complex forward transmission amplitude for probe helicity $\eta$, and $\kappa_{\rm ext}$ is the coupling rate to the measured port. In Fig.~\ref{fig:spectroscopy}(d), we insert the exact matter susceptibility into Eq.~\eqref{eq:inputoutput} to obtain the illustrative transmitted intensities $|S_{21}^{(\eta)}|^2$. This input--output step only shows how the matter threshold and helicity contrast enter a cavity measurement. The sum rule and topological results do not depend on the chosen probe parameters.
The helicity difference rejects reciprocal backgrounds and inherits both the threshold and the dichroic spectral weight. Fig.~\ref{fig:spectroscopy} shows the exact spectrum, cumulative sum rule, mass scan, and representative cavity readout. The unweighted integral is a chirality sum rule. It is not by itself the quantized CD response discussed for filled topological bands~\cite{Tran2017}.
Further discussion of the threshold, cumulative sum rule, and device-dependent transmission line shape in Fig.~\ref{fig:spectroscopy} is given in the Supplemental Material.

\begin{figure}[t]
\centering
\includegraphics[width=0.99\linewidth]{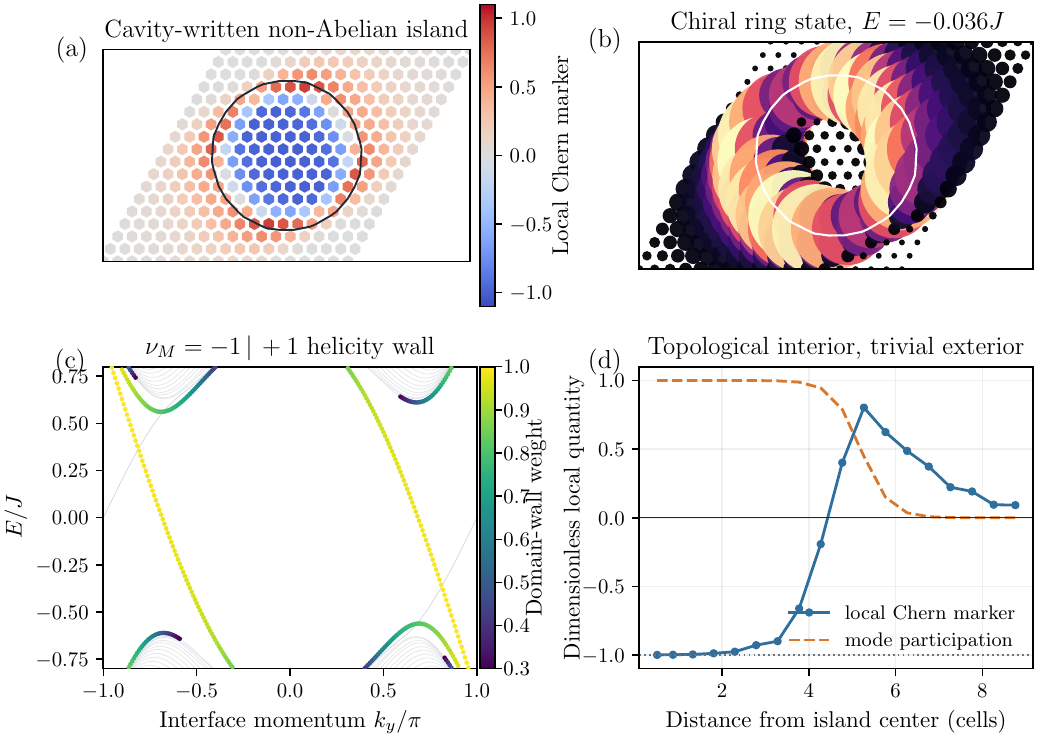}
\caption{\label{fig:realspace}Spatial programming of Ising topology. (a) A localized mass profile inside a trivial anisotropic background creates a non-Abelian island whose local Chern marker approaches $-1$. (b) The lowest-magnitude finite-size state is a chiral ring localized at the island boundary. (c) A smooth $\nu_M=-1|+1$ helicity wall supports two co-propagating Majorana modes. (d) The marker follows the cavity participation profile from topological interior to trivial exterior.}
\end{figure}

The participation theorem makes a localized implementation a natural physical target. For a smooth cavity-written island inside a trivial anisotropic exterior, the local Chern marker~\cite{BiancoResta2011} reaches $-0.977\pm0.025$ in the interior, while the lowest-magnitude state forms the chiral boundary ring required by bulk--edge correspondence. A helicity wall between $\kappa>0$ and $\kappa<0$ has $|\Delta\nu_M|=2$ and therefore carries two co-propagating Majorana modes. The construction and finite-size details are given in the Supplemental Material, while the spatial crossover is shown in Fig.~\ref{fig:realspace}. Arrays of localized chiral modes can thus write islands, links, and domain walls without requiring a uniform macroscopic cavity mass.

The four panels of Fig.~\ref{fig:realspace} distinguish a topological island from a merely inhomogeneous gap. Panel (a) exhibits a broad $C\simeq-1$ plateau away from both the sample edge and cavity boundary. Panel (b) places the lowest-magnitude state on that boundary rather than in the island interior. Panel (c) removes the trivial exterior and resolves the two interface branches implied by $|\Delta\nu_M|=2$, while panel (d) shows that the marker follows the smooth participation profile rather than an imposed lattice cut. The mode waist sets the island footprint, its helicity fixes the propagation direction, and overlapping waists can define reconfigurable neutral channels. The finite ring level spacing imposes a lower size constraint of order $\hbar v/L$.

A uniform real-time helicity reversal cannot be adiabatic in the thermodynamic limit. Near each valley, with $\rho_{x,y,z}$ the two-band pseudospin, $H_{\bm q}(t)=v(q_x\rho_x+q_y\rho_y)+m(t)\rho_z$ with $m=t/\tau_Q$ gives the Landau--Zener excitation density
$n_{\rm ex}=\frac{N_D}{4\pi^2v^2\tau_Q}$, here $N_D=2$,
a two-dimensional quantum Kibble--Zurek law~\cite{Landau1932,Zener1932,Kibble1976,Zurek1985,Dziarmaga2010,Polkovnikov2011}.
Moving a preexisting wall instead localizes the unavoidable gap closing at the interface while leaving the distant bulk gapped (Supplemental Material and Fig.~\ref{fig:switching}).

A ground-state vacuum Lamb shift and a resonant probe are distinct. Let $\delta E_\eta^{(2)}$ ($\eta=\pm1$) denote the second-order correction to the matter ground-state energy from virtual exchange with a cavity photon in the vacuum (see Supplementary Materials for details), for cavity helicity $\eta$. Its helicity-odd part is
\begin{equation}
\delta E_+^{(2)}-\delta E_-^{(2)}
=-g^2\int_0^\infty\mathrm{d}\Omega
\frac{I_+(\Omega)-I_-(\Omega)}{\hbar(\omega_c+\Omega)}.
\label{eq:continuumkernel}
\end{equation}
This difference is the helicity-odd vacuum Lamb shift: it is a virtual ground-state energy renormalization whose low-frequency, large-$\omega_c$ limit produces the effective chiral mass. The helicity-even part shifts both helicities equally and is omitted here. Eq.~\eqref{eq:continuumkernel} is not an absorption rate or a resonant transition energy.
The denominator is a \emph{sum} frequency, so there is no ground-state resonance at $\omega_c=\Omega$. At large $\omega_c$, Eq.~\eqref{eq:continuumkernel} recovers the commutator through Eq.~\eqref{eq:sumrule}. At moderate frequency the entire Majorana continuum contributes. For the illustrated mass, the leading high-frequency approximation still differs from the full kernel by about $15\%$ at $\omega_c=10J$. A passive linewidth weakly modifies this virtual kernel rather than defining a spontaneous ground-state decay rate, whereas the weak probe in Eq.~\eqref{eq:inputoutput} is resonant at matter transitions. The full continuum and linewidth analysis is shown in the Supplemental Material and Fig.~\ref{fig:continuum}.
For vacuum stabilization, cavity loss therefore primarily renormalizes the virtual kernel. The operational scale is $\Dop=\min(\Delta_\psi,\Delta_\sigma,\Delta_{\rm bulk})$ and should exceed $k_BT$, intrinsic spin broadening, and the finite-island level spacing $\hbar v/L$.

Taken together, these results illustrate the physical meaning of cavity-induced non-Abelian order. The algebraic mechanism is universal: a chiral vacuum converts noncommuting projected polarization components into a time-reversal-odd interaction. The honeycomb bond polarization makes this interaction the Kitaev mass exactly. Mode normalization then decides the spatial outcome. One extended normalized mode is not a thermodynamic bulk field. A localized mode writes a non-Abelian island, and a finite mode density can generate a bulk mass. The same polarization algebra supplies an exact dichroic measurement, while spatial helicity profiles create interfaces with a topologically fixed number of Majorana channels.

These results suggest a hierarchy of experimental goals. The most direct single-cavity demonstration is a local topological droplet with a helicity-dependent ring spectrum and integrated CD. A cavity array or multimode photonic structure is the natural route to a bulk phase. Uniform helicity reversal should be viewed as a controlled topological quench rather than an adiabatic switch. Domain-wall motion offers a spatially localized reconfiguration protocol. Finally, vacuum stabilization and probe absorption must be modeled with different denominators. The vacuum term samples the entire matter continuum through $\omega_c+\Omega_n$, where $\Omega_n=(E_n-E_0)/\hbar>0$ is a discrete matter-transition frequency; probe absorption is resonant at $\omega=\Omega_n$. This equilibrium mechanism is consequently distinct from Floquet topological engineering by populated circular drives~\cite{OkaAoki2009,Lindner2011}.

The exact gapped phase is stable to sufficiently weak local perturbations~\cite{BravyiHastingsMichalakis2010}.
The remaining material problem is quantitative but now well posed: compute the full projected-polarization tensor and normalized chiral electromagnetic kernel, then derive every same-order spin interaction. These steps determine whether a candidate realizes a cavity-written local Ising region or an intensive bulk non-Abelian chiral spin liquid. The interacting thermodynamic demonstration is left to future work.

\begin{acknowledgments}
\hbadness=2000
\begin{sloppypar}
We acknowledge support from the US DOE, Office of Science, Basic Energy Sciences, Chemical Sciences, Geosciences, and Biosciences Division under Triad National Security, LLC (``Triad'') contract Grant 89233218CNA000001 (FWP: LANLECF7).
This research used computational resources provided by the Institutional Computing (IC) Program at Los Alamos National Laboratory (LANL). LANL is operated by Triad National Security, LLC, for the National Nuclear Security Administration of the US Department of Energy (Contract No. 89233218CNA000001).
\end{sloppypar}
\end{acknowledgments}

\bibliography{references}

\clearpage
\begin{widetext}

 \appendix
 \begin{center}
 \large\bf Supplemental Material for ``\mytitle"
 \end{center}

\def\CGBSupplementBodyOnly{1}
\ifdefined\CGBSupplementBodyOnly
\newcommand{\CGBStandaloneEnd}{}
\else
\documentclass[aps,prb,onecolumn,superscriptaddress,nofootinbib,floatfix]{revtex4-2}
\usepackage{amsmath,amssymb,bm,graphicx,booktabs,array,microtype}
\usepackage[colorlinks=true,citecolor=blue,linkcolor=blue,urlcolor=blue]{hyperref}
\newcommand{\kcav}{\kappa_{\mathrm{cav}}}
\newcommand{\kph}{\kappa_{\mathrm{ph}}}
\newcommand{\Othree}{O_{3}}
\newcommand{\Pzero}{\mathcal P_0}
\newcommand{\Qzero}{\mathcal Q_0}
\newcommand{\Dop}{\Delta_{\mathrm{op}}}
\begin{document}

\title{Supplemental Material for ``Chiral Cavity Vacua as Local Generators and Probes of Non-Abelian Kitaev Topology''}
\author{Yu Zhang}
\affiliation{Theoretical Division, Los Alamos National Laboratory, Los Alamos, New Mexico 87545, USA}
\date{\today}
\maketitle

\newcommand{\CGBStandaloneEnd}{\end{document}}
\fi

\setcounter{figure}{0}
\renewcommand{\thefigure}{S\arabic{figure}}
\setcounter{equation}{0}
\renewcommand{\theequation}{S\arabic{equation}}

\section{Scope, hierarchy of projections, and claim boundaries}

The analysis contains two distinct downfoldings. First, a microscopic spin-orbit-coupled Mott insulator is projected from charge-transfer states onto one low-energy Kramers doublet per magnetic site. This step generates both a spin Hamiltonian $H_s$ and a projected electric-polarization operator $\bm P^{\rm spin}$. Second, virtual cavity photons are eliminated within the spin manifold. The first step is material dependent. The second produces the operator identities used in the main text.

A schematic microscopic Hamiltonian is
\begin{equation}
H_{\rm mic}=H_{\rm loc}+T+H_{\rm ph}-\bm E\cdot\bm D+H_{\rm DSE},
\end{equation}
where $H_{\rm loc}$ contains local interactions, crystal fields, and spin-orbit coupling, $T$ contains metal-ligand and intersite hopping, $\bm D$ is the microscopic dipole, and $H_{\rm DSE}$ is the dipole self-energy. A controlled hierarchy for the compact analytic treatment is
\begin{equation}
\|T\|,\hbar\omega_c,\|H_s\|\ll\Delta_{\rm ch},
\qquad
\|gP_a\|\ll\hbar\omega_c,
\label{eq:hierarchy}
\end{equation}
with $\Delta_{\rm ch}$ the lowest eliminated charge-transfer scale. The first inequality controls the electronic Schrieffer--Wolff transformation. The second controls photon-vacuum elimination. The additional high-frequency expansion in Sec.~\ref{sec:vacsw} requires $\hbar\omega_c$ to exceed the spin-energy transfers sampled by the projected polarization. The exact resolvent remains applicable when that last expansion is not accurate.

The numerical calculations establish the following statements without an interacting thermodynamic simulation: (i) a gyrotropic vacuum generates the helicity-odd commutator $\mathrm{i}[P_x,P_y]$; (ii) the ideal Kitaev bond polarization converts this commutator into the oriented three-spin mass; (iii) the Kitaev $B$ phase becomes an Ising chiral spin liquid with a helicity-reversed Majorana Chern number and edge propagation; (iv) a normalized single extended mode produces a local mass that decreases with lateral area; (v) a localized mode produces a finite topological island; and (vi) the integrated circular dichroism obeys the exact sum rule derived below.

\section{Gauge-consistent projected light--matter Hamiltonian}
\label{sec:gauge}

After projection to the spin manifold, a localized circular mode is written
\begin{align}
H={}&H_s+\hbar\omega_c a^\dagger a
+g\left(aQ_\eta^\dagger+a^\dagger Q_\eta\right)+H_{\rm DSE},\nonumber\\
Q_\eta={}&\frac{P_x+\mathrm{i}\eta P_y}{\sqrt2},\qquad \eta=\pm1.
\label{eq:projectedH}
\end{align}
Here $P_x$ and $P_y$ are Hermitian operators in the low-energy spin Hilbert space. The dipole self-energy must be retained because omitting it after a matter truncation can violate gauge consistency and alter the helicity-even interaction sector \cite{DeBernardis2018,Schaefer2020,Dmytruk2021,Ashida2021}. The helicity-odd result below follows from the antisymmetric operator algebra and is not canceled by a time-reversal-even self-polarization term.

\section{Exact vacuum Schrieffer--Wolff transformation}
\label{sec:vacsw}

Let $P_{\rm vac}=|0\rangle\langle0|$ project onto the photon vacuum. A Schrieffer--Wolff transformation~\cite{SchriefferWolff1966,BravyiDiVincenzoLoss2011} gives, to second order in $g$, the effective spin Hamiltonian at a spin-sector reference energy $E$,
\begin{equation}
\delta H_\eta^{(2)}=-g^2Q_\eta^\dagger
\frac{1}{\hbar\omega_c+H_s-E}Q_\eta.
\label{eq:exactres}
\end{equation}
Expanding the resolvent in $(H_s-E)/(\hbar\omega_c)$ gives
\begin{align}
\delta H_\eta^{(2)}={}&-\frac{g^2}{\hbar\omega_c}Q_\eta^\dagger Q_\eta
+\frac{g^2}{(\hbar\omega_c)^2}Q_\eta^\dagger(H_s-E)Q_\eta+\cdots,\\
Q_\eta^\dagger Q_\eta={}&\frac12(P_x^2+P_y^2)+\frac{\eta}{2}\mathrm{i}[P_x,P_y].
\end{align}
Therefore
\begin{equation}
\boxed{
\delta H_\eta^{(2)}=-\lambda(P_x^2+P_y^2)
-\eta\lambda\mathrm{i}[P_x,P_y]+O(\omega_c^{-2}),
\quad
\lambda=\frac{g^2}{2\hbar\omega_c}.}
\label{eq:SWidentity}
\end{equation}
Because electric polarization is time-reversal even while time reversal complex conjugates $\mathrm{i}$, the commutator term is time-reversal odd. It reverses under $\eta\rightarrow-\eta$, whereas the symmetric term does not.

The next term,
\begin{equation}
\delta H_\eta^{(2,1)}=
\frac{g^2}{(\hbar\omega_c)^2}Q_\eta^\dagger(H_s-E)Q_\eta,
\end{equation}
generates energy-dependent and longer-range spin interactions. Near the spin bandwidth one should use Eq.~\eqref{eq:exactres}, or its spectral representation in Sec.~\ref{sec:continuum}, rather than the leading commutator alone.

We diagonalize three spins coupled to a truncated photon mode. The operator-norm difference between the exact helicity-odd second-order block and the commutator expression is $4.46\times10^{-16}$. After including the leading $O(g^2)$ correction, the residual low-energy spectral error has fitted power $3.991$, consistent with the expected fourth-order scaling.

\section{Normalized modes and the thermodynamic participation theorem}
\label{sec:participation}

\subsection{Local overlap structure}

Resolve the projected polarization into local bond operators,
\begin{equation}
P_a[u]=\sum_b u_bP_{a,b},
\qquad
\sum_b|u_b|^2=1.
\label{eq:Pu}
\end{equation}
The normalization is inherited from the electromagnetic energy normalization of one mode. The odd operator is
\begin{equation}
\mathrm{i}[P_x[u],P_y[u]]=
\sum_{bb'}u_b^*u_{b'}\mathrm{i}[P_{x,b},P_{y,b'}].
\label{eq:commu}
\end{equation}
For local spin operators, the commutator vanishes unless $b$ and $b'$ overlap. Let ${\cal N}(j)$ denote the finite set of ordered overlapping bond pairs that generate a three-spin term centered near $j$. Its coefficient has the form
\begin{equation}
\kappa_j\propto\sum_{(b,b')\in{\cal N}(j)}u_b^*u_{b'}c_{bb'},
\label{eq:localmass}
\end{equation}
where $c_{bb'}$ is independent of total sample area for a fixed local material Hamiltonian.

For a laterally extended normalized mode over $N_b\propto A$ bonds, $|u_b|\sim N_b^{-1/2}$. The number of terms in ${\cal N}(j)$ is $O(1)$, so
\begin{equation}
\boxed{\kappa_j^{\rm extended}=O(A^{-1}).}
\label{eq:extscaling}
\end{equation}
The sum over all sites is $O(1)$, but the local mass and the thermodynamic bulk gap vanish. This scaling distinguishes the helicity-odd local commutator from an unnormalized square of the collective polarization, whose cross terms can appear to scale superextensively if mode normalization is omitted.

We quantify mode localization using the participation measure
\begin{equation}
N_{\rm eff}=\frac{(\sum_b|u_b|^2)^2}{\sum_b|u_b|^4}
=\frac{1}{\sum_b|u_b|^4}.
\end{equation}
For a mode localized over a fixed region, $N_{\rm eff}$ remains finite as the surrounding sample grows. Equation~\eqref{eq:localmass} then gives a finite mass within the mode volume and a vanishing mass outside. A finite-density array of localized modes gives an intensive bulk mass because the number of contributing modes scales with area.

\subsection{Multimode and Green-tensor formulation}

For modes $\lambda$ with spatial profiles $u_{\lambda b}$, the leading local mass is
\begin{equation}
\kappa_j=\sum_\lambda
\frac{g_\lambda^2\eta_\lambda}{2\hbar\omega_\lambda}
\sum_{(b,b')\in{\cal N}(j)}u_{\lambda b}^*u_{\lambda b'}c_{bb'}.
\label{eq:multimode}
\end{equation}
In a continuum environment, the mode sum is replaced by the antisymmetric part of the positive-frequency electric-field correlation tensor. A compact local design quantity is the chiral electric spectral density
\begin{equation}
\rho_{\rm ch}^{\rm loc}(\bm r,\omega)=
\sum_\lambda\left(
|\bm e_+\cdot\bm E_\lambda(\bm r)|^2-
|\bm e_-\cdot\bm E_\lambda(\bm r)|^2
\right)\delta(\omega-\omega_\lambda).
\label{eq:localchiralLDOS}
\end{equation}
A reciprocal environment with degenerate opposite-helicity modes gives zero after the mode sum. A bulk mass requires that the spatially contracted chiral spectral density, not merely the total density of states or a collective splitting, remain finite in the thermodynamic limit.

\section{Projected spin polarization and exact honeycomb commutator}
\label{sec:polarization}

\subsection{Electronic downfolding}

Let $\Pzero$ project onto one Kramers doublet per site and $\Qzero=1-\Pzero$ contain charge excitations. A canonical transformation $e^S$, with $S=-S^\dagger$, gives
\begin{equation}
P_a^{\rm spin}=\Pzero e^SD_ae^{-S}\Pzero
=\Pzero\left(D_a+[S,D_a]+\frac12[S,[S,D_a]]+\cdots\right)\Pzero.
\end{equation}
For a Mott insulator, $S=O(T/\Delta_{\rm ch})$. If the projected bare dipole is a constant and the first-order term changes charge, the leading spin-dependent contribution is second order. A useful resolvent expression is
\begin{align}
P_a^{\rm spin}={}&\Pzero T R D_a R T\Pzero
-\frac12\Pzero\{TR^2T,D_a\}\Pzero+\cdots,\\
R={}&\Qzero(E_0-H_{\rm loc})^{-1}\Qzero.
\label{eq:Presolvent}
\end{align}
On a bond, the most general bilinear component can be parameterized as
\begin{equation}
P_{ij}^a=\sum_{\mu\nu}\Pi_{ij}^{a;\mu\nu}
\sigma_i^\mu\sigma_j^\nu+\bm p_{ij}^{(1)}\cdot(\bm\sigma_i+\bm\sigma_j)+p_{ij}^{(0)}.
\label{eq:Pi}
\end{equation}
The ideal Kitaev component is
\begin{equation}
\bm P^{\rm spin}=p\sum_{\langle ij\rangle_\gamma}
-\bm\delta_\gamma\sigma_i^\gamma\sigma_j^\gamma+\delta\bm P.
\label{eq:idealP}
\end{equation}
The coefficient $p$ scales parametrically as $ea(t/\Delta_{\rm ch})^2$ times spin-orbit and inversion-asymmetry form factors. Bond inversion can force the longitudinal component to vanish. An interface, substrate, gate, asymmetric cavity stack, unequal ligands, or intrinsic odd-parity charge transfer can remove this cancellation. A quantitative material calculation must extract the full tensor $\Pi_{ij}^{a;\mu\nu}$, not only $p$.

\subsection{Exact commutator algebra}

Choose directed bond vectors
\begin{equation}
\bm\delta_x=(1,0),\quad
\bm\delta_y=(-1/2,\sqrt3/2),\quad
\bm\delta_z=(-1/2,-\sqrt3/2),
\end{equation}
and define
\begin{equation}
B_\gamma=\sum_{\langle ij\rangle_\gamma}\sigma_i^\gamma\sigma_j^\gamma.
\end{equation}
Then
\begin{equation}
P_x=p(B_x-\tfrac12B_y-\tfrac12B_z),
\qquad
P_y=\tfrac{\sqrt3p}{2}(B_y-B_z).
\end{equation}
Disjoint bonds commute. Two distinct Kitaev bonds sharing a site $j$ give, for example,
\begin{equation}
\mathrm{i}[\sigma_i^x\sigma_j^x,\sigma_j^y\sigma_k^y]
=-2\sigma_i^x\sigma_j^z\sigma_k^y.
\end{equation}
Collecting all ordered adjacent-bond pairs yields
\begin{align}
\mathrm{i}[P_x,P_y]&=-\sqrt3p^2\Othree,\label{eq:commmass}\\
\Othree&=-\frac{\mathrm{i}}{2}\left([B_x,B_y]+[B_y,B_z]+[B_z,B_x]\right).
\end{align}
Thus
\begin{equation}
H_{\rm vac}^{\rm odd}=\kcav\Othree,
\qquad
\kcav=\eta\frac{\sqrt3g^2p^2}{2\hbar\omega_c}.
\label{eq:masscoeff}
\end{equation}
The symmetric operator $P_x^2+P_y^2$ contains exchange renormalizations, long-range bond-bond interactions for an extended mode, and local four-spin terms for a finite-range kernel. The dipole self-energy and the same electronic downfolding also modify that even sector. It cannot be dropped in a physical material trajectory.

\section{Exact Majorana solution, gap, and topology}
\label{sec:majorana}

In the flux-free sector, $\sigma_i^\gamma=\mathrm{i} b_i^\gamma c_i$ maps the model to nearest-neighbor and oriented next-nearest-neighbor Majorana hopping. In a two-site unit cell, the two-band sublattice pseudospin is denoted by $\rho_{x,y,z}$:
\begin{equation}
\mathcal H(\bm k)=d_x\rho_x+d_y\rho_y+d_z\rho_z,
\end{equation}
with
\begin{align}
f(\bm k)=d_x+\mathrm{i} d_y&=J_z+J_xe^{\mathrm{i} k_1}+J_ye^{\mathrm{i} k_2},\\
d_z(\bm k)&=2\kcav[\sin k_1-\sin k_2+\sin(k_2-k_1)].
\end{align}
The bands are $E_\pm=\pm\sqrt{|f|^2+d_z^2}$.

At $\kcav=0$, zeros of $f$ exist exactly when $|J_x|$, $|J_y|$, and $|J_z|$ satisfy the triangle inequalities. For $J_x=J_y=J>0$ and $r=J_z/J$, the Dirac points can be chosen as
\begin{equation}
k_1^*=\pi+\theta,\qquad k_2^*=\pi-\theta,
\qquad \cos\theta=r/2,
\end{equation}
for $0\le r\le2$. At these points,
\begin{equation}
|d_z(\bm k^*)|=4|\kcav|\left(1+\frac r2\right)
\sqrt{1-\frac{r^2}{4}},
\end{equation}
so the direct band gap is
\begin{equation}
\Delta_M=8|\kcav|\left(1+\frac r2\right)
\sqrt{1-\frac{r^2}{4}}.
\label{eq:gapani}
\end{equation}
For $r=1$, $\Delta_M=6\sqrt3|\kcav|$. At $r=2$, the cones merge at a zero of the chiral form factor. For $r>2$, the unperturbed system is in the Abelian $A$ phase and an infinitesimal chiral term does not change its topological index.

For a two-band model, the occupied-band Chern number is
\begin{equation}
C_{\rm occ}=\frac{1}{4\pi}\int_{\rm BZ}\mathrm{d}^2k
\hat{\bm d}\cdot(\partial_{k_1}\hat{\bm d}\times\partial_{k_2}\hat{\bm d}).
\end{equation}
The two valleys have opposite kinetic winding and opposite $d_z$, so their half-integer contributions add. In the reciprocal orientation,
\begin{equation}
C_{\rm occ}=-\operatorname{sgn}(J_xJ_yJ_z\kcav).
\end{equation}
Reversing the real-space orientation reverses this bookkeeping sign, but the relation $\nu_M(\eta)=-\nu_M(-\eta)$ is invariant. We evaluate the invariant using the Fukui--Hatsugai--Suzuki lattice discretization \cite{Fukui2005}.

For a strip, periodic momentum is retained along one Bravais direction while the transverse direction is open. Spatially resolved eigenvector weight separates bulk states from the two physical boundaries. Opposite signs of $\kcav$ exchange the edge slopes while leaving every bulk eigenvalue invariant.

The panels of Fig.~2 of the main text isolate complementary parts of these results. Panel (a) shows that the chiral term changes the eigenvectors and opens the Dirac gap without shifting the cone positions in the weak-mass regime. Panel (b) confirms the predicted linear slope $6\sqrt3$ and therefore tests the normalization of $\Othree$, not only the presence of a gap. Panels (c) and (d) show why anisotropy matters: the mass is topological only while the parent couplings satisfy the $B$-phase triangle condition, and its efficiency vanishes continuously when the valleys merge at $J_z/J=2$. Thus increasing the vacuum coupling cannot compensate for a parent state driven beyond the $B$-phase boundary. The cavity is most effective when the three Kitaev exchanges are balanced. Finally, panels (e) and (f) provide the boundary counterpart of the bulk invariant. The bulk continua coincide for $\pm\kcav$, whereas the in-gap branch reverses velocity, making helicity reversal a direct switch of neutral edge propagation rather than a change in the bulk density of states. Taken together, the panels distinguish a genuine topological mass from a generic spectral gap: it must gap the parent Dirac cones, carry the quantized bulk index, and produce the corresponding helicity-reversible edge branch.

Representative exact checks are
\begin{center}
\begin{tabular}{ccccc}
\toprule
$J_x$ & $J_y$ & $J_z$ & $\kcav$ & $(\Delta_M,C_{\rm occ})$\\
\midrule
1 & 1 & 1 & $+0.06$ & $(0.623538,-1)$\\
1 & 1 & 1 & $-0.06$ & $(0.623538,+1)$\\
1 & 1 & 2.4 & $+0.06$ & $(\text{finite},0)$\\
\bottomrule
\end{tabular}
\end{center}

\section{Circular absorption and the exact dichroism sum rule}
\label{sec:spectroscopy}

\subsection{Completeness identity}

Define
\begin{equation}
I_\eta(\omega)=\sum_{n\ne0}|\langle n|Q_\eta|0\rangle|^2
\delta[\omega-(E_n-E_0)/\hbar].
\end{equation}
Integrating over positive frequency and inserting completeness gives
\begin{align}
\int_0^\infty\mathrm{d}\omega I_\eta(\omega)
={}&\langle0|Q_\eta^\dagger Q_\eta|0\rangle
-|\langle0|Q_\eta|0\rangle|^2.
\end{align}
The elastic terms are equal for $\eta=\pm1$ because $P_x$ and $P_y$ are Hermitian. Subtracting the two helicities yields
\begin{equation}
\boxed{
\int_0^\infty\mathrm{d}\omega [I_+(\omega)-I_-(\omega)]
=\langle\mathrm{i}[P_x,P_y]\rangle.}
\label{eq:sumrule_sm}
\end{equation}
Using Eq.~\eqref{eq:commmass},
\begin{equation}
\int_0^\infty\mathrm{d}\omega [I_+-I_-]
=-\sqrt3p^2\langle\Othree\rangle
=-\sqrt3p^2\frac{\partial E_0}{\partial\kcav}.
\label{eq:sumO3}
\end{equation}
The sum rule is exact for any interacting spin Hamiltonian with the stated definition of $Q_\eta$. It measures chirality. It is not automatically a quantized Chern-number sum rule.

\subsection{Exact Kitaev spectrum}

For the ideal bond polarization, the flux-conserving Bloch vertices used in the calculation are
\begin{align}
g_x(\bm k)&=p\left(e^{\mathrm{i} k_1}-\tfrac12e^{\mathrm{i} k_2}-\tfrac12\right),\\
g_y(\bm k)&=\frac{\sqrt3p}{2}(e^{\mathrm{i} k_2}-1),\\
V_a(\bm k)&=\begin{pmatrix}0&g_a^*\\g_a&0\end{pmatrix},
\qquad V_\eta=\frac{V_x+\mathrm{i}\eta V_y}{\sqrt2}.
\end{align}
If $|u_-(\bm k)\rangle$ and $|u_+(\bm k)\rangle$ are occupied and empty Bloch eigenvectors, then
\begin{equation}
I_\eta(\omega)=\frac1{N_k}\sum_{\bm k}
|\langle u_+|V_\eta|u_-\rangle|^2
\delta[\omega-2E(\bm k)].
\end{equation}
The threshold is the itinerant Majorana gap. For $J_x=J_y=J_z=1$, $\kcav=0.06J$, and $p=1$, a $181\times181$ momentum grid gives
\begin{align}
\omega_{\rm th}&=0.62383834339,\\
\int(I_+-I_-)\mathrm{d}\omega&=0.403489470134,\\
-\sqrt3\langle\Othree\rangle&=0.403489470134,
\end{align}
with absolute sum-rule error $5.55\times10^{-17}$. The slight difference between $\omega_{\rm th}$ and $6\sqrt3(0.06J)=0.623538J$ is the finite momentum-grid resolution.

The positive-frequency retarded susceptibility is evaluated as
\begin{equation}
\chi_\eta^R(\omega)=\sum_n w_{n\eta}
\left[\frac{1}{\omega-\Omega_n+\mathrm{i}\gamma}
-\frac{1}{\omega+\Omega_n+\mathrm{i}\gamma}\right].
\end{equation}
For a weak probe cavity, a schematic input--output denominator is
\begin{equation}
D_\eta(\omega)=\mathrm{i}(\omega_c-\omega)+\frac{\kph}{2}
+\mathrm{i} g_{\rm pr}^2\chi_\eta^R(\omega),
\qquad
S_{21}^{(\eta)}=1-\frac{\kappa_{\rm ext}}{D_\eta}.
\end{equation}
The example in the main text uses dimensionless probe parameters only to display the exact matter threshold and helicity contrast, it is not a fit to a specific device.

Figure~3(a) of the main text first separates two experimental observables: the common absorption onset locates the itinerant Majorana gap, while the difference between helicities carries the sign-sensitive chiral information. Because the differential spectrum changes sign across the continuum, a single-frequency contrast does not measure the mass generator. The cumulative curve in panel (b) demonstrates why spectral integration is essential: only after the full two-Majorana continuum is included does it saturate at $\langle\mathrm{i}[P_x,P_y]\rangle$. Panel (c) repeats the comparison for both mass signs and a range of magnitudes, showing that the equality is an operator identity rather than an accidental agreement at one parameter point. Panel (d) translates the same susceptibility into a practical weak-probe observable. Its role is to illustrate how the threshold and helicity contrast enter transmission, while the absolute line shape remains device dependent.

\section{Equilibrium vacuum continuum versus resonant probe response}
\label{sec:continuum}

\begin{figure}[t]
\centering
\includegraphics[width=0.82\linewidth]{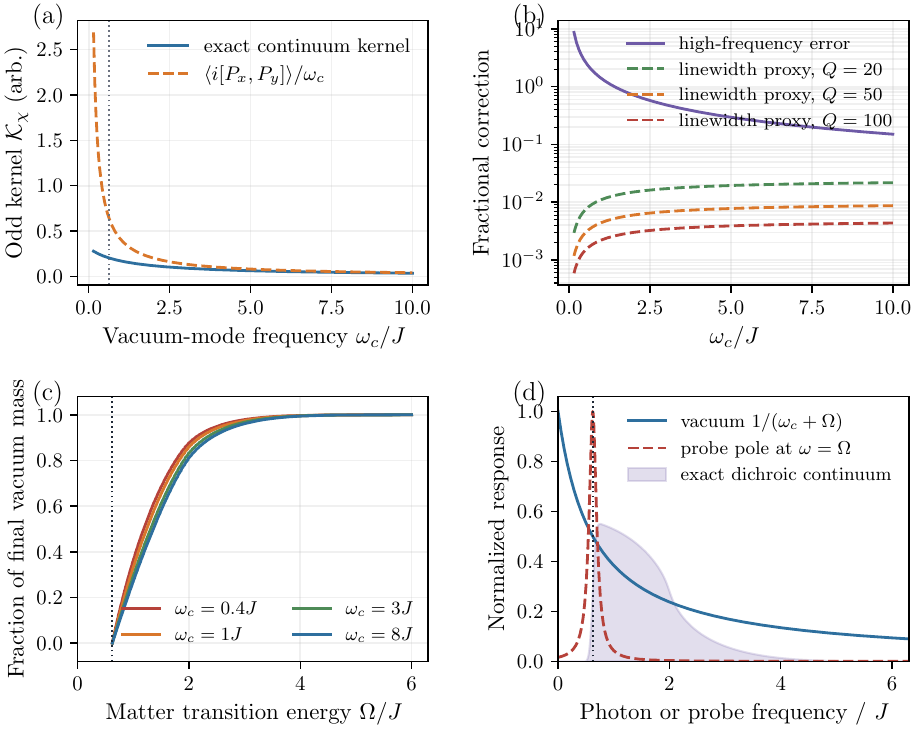}
\caption{\label{fig:continuum}Exact continuum control of the vacuum mass. (a) The equilibrium helicity-odd kernel uses sum-frequency denominators and approaches the commutator result only at high cavity frequency. (b) High-frequency and finite-linewidth corrections are distinct. (c) A broad range of two-Majorana transitions contributes to the vacuum mass. (d) The equilibrium factor $1/(\omega_c+\Omega)$ has no resonance at the matter gap, whereas the weak-probe response has a pole at $\omega=\Omega$.}
\end{figure}

For the matter ground state, the spectral representation of Eq.~\eqref{eq:exactres} is
\begin{equation}
\delta E_\eta^{(2)}=-g^2\sum_n
\frac{|\langle n|Q_\eta|0\rangle|^2}
{\hbar\omega_c+E_n-E_0}.
\label{eq:spectralshift}
\end{equation}
The equilibrium vacuum denominator contains a sum of positive frequencies. It does not have a pole at $\omega_c=(E_n-E_0)/\hbar$. Define the odd continuum kernel
\begin{equation}
{\cal K}_\chi(\omega_c)=\sum_n
\frac{|\langle n|Q_+|0\rangle|^2-|\langle n|Q_-|0\rangle|^2}
{\omega_c+\Omega_n}.
\label{eq:Kchi}
\end{equation}
At large $\omega_c$,
\begin{equation}
{\cal K}_\chi(\omega_c)=
\frac{\langle\mathrm{i}[P_x,P_y]\rangle}{\omega_c}
-\frac{1}{\omega_c^2}\sum_n\Omega_n(w_{n+}-w_{n-})+\cdots.
\label{eq:hfexp}
\end{equation}
For the exact spectrum used in Fig.~\ref{fig:continuum}, the leading commutator approximation is still $15.0\%$ away from the full continuum at $\omega_c=10J$, illustrating that the first frequency moment can be important even when the operator structure is correct.

A phenomenological replacement
\begin{equation}
\frac{1}{\omega_c+\Omega_n}
\longrightarrow
\frac{1}{\omega_c+\Omega_n-\mathrm{i}\kph/2}
\end{equation}
can estimate sensitivity to a finite photonic linewidth. The imaginary part of this expression is an engineering propagator proxy. It is not interpreted here as a spontaneous decay rate of the exact ground state of the complete matter--photon--bath system. By contrast, weak-probe absorption is resonant through denominators $\omega-\Omega_n+\mathrm{i}0^+$. This distinction is why a single-pole ratio such as $2|\Delta|/\kph$ is not used as a universal criterion for the equilibrium vacuum Lamb shift.

Figure~\ref{fig:continuum} displays the full comparison. The high-frequency curve tests the domain of the leading commutator expansion, while the cumulative curves show that a broad interval of the two-Majorana continuum contributes at moderate $\omega_c$. The finite-linewidth curves are propagator proxies, not ground-state decay rates. For vacuum stabilization, cavity loss therefore primarily renormalizes the virtual kernel. The operational topological scale should satisfy
\begin{equation}
\Dop=\min(\Delta_\psi,\Delta_\sigma,\Delta_{\rm bulk})
>\max(k_BT,\hbar\Gamma_{\rm spin},\hbar v/L),
\end{equation}
while a spectroscopic measurement separately requires adequate probe resolution and coupling.

\section{Real-space topological island and helicity wall}
\label{sec:realspace}

\subsection{Island Hamiltonian and local Chern marker}

The island calculation uses an open $18\times18$ array of two-site cells. A smooth participation profile
\begin{equation}
w(\bm r)=\frac12\left[1-\tanh\frac{|\bm r-\bm r_0|-R}{\ell}\right]
\end{equation}
sets
\begin{equation}
\kappa(\bm r)=0.14J w(\bm r),
\qquad
J_z(\bm r)=2.55J+(J-2.55J)w(\bm r).
\end{equation}
Thus the interior is a massive Kitaev $B$ phase and the exterior is a strongly anisotropic trivial $A$ phase. Nearest-neighbor and chiral next-nearest-neighbor hoppings are averaged across spatial links.

For a finite single-particle Hamiltonian with negative-energy projector $P$ and $Q=1-P$, the cell-resolved local Chern marker is evaluated from \cite{BiancoResta2011}
\begin{equation}
C(\bm r)=-\frac{4\pi}{A_{\rm cell}}
\operatorname{Im}\langle\bm r|P\hat xQ\hat yP|\bm r\rangle,
\label{eq:marker}
\end{equation}
with the two sublattice contributions summed in each cell. The matrix dimension is 648. The central mean marker is $-0.9774547$ with standard deviation $0.02456$. The selected exterior annulus has mean $+0.1129$, a finite-open-boundary contamination rather than a bulk topological value. The eigenstate nearest zero has energy $-0.03557J$ and unit total probability. Its weight is concentrated on the island boundary.

\subsection{Helicity domain wall}

For a strip open in $x$ and periodic in $y$, the profile
\begin{equation}
\kappa(x)=\kappa_0\tanh[(x-x_0)/\ell]
\end{equation}
interpolates between $\nu_M=-1$ and $\nu_M=+1$. Bulk--interface correspondence requires
\begin{equation}
N_{\rm chiral\ Majorana}=|\Delta\nu_M|=2.
\end{equation}
The strip spectrum is diagonalized for 58 transverse cells and 221 momenta. Interface-localized weight identifies two co-propagating Majorana branches, equivalent to one neutral complex chiral channel. 

\section{Uniform helicity reversal and Kibble--Zurek scaling}
\label{sec:switching}

Near one valley, the continuum Hamiltonian is
\begin{equation}
H_{\bm q}(t)=v(q_x\rho_x+q_y\rho_y)+m(t)\rho_z,
\qquad m(t)=t/\tau_Q.
\end{equation}
After a basis rotation this is a Landau--Zener crossing with transition probability
\begin{equation}
p_{\bm q}=\exp(-\pi v^2q^2\tau_Q).
\end{equation}
For one Dirac cone,
\begin{align}
n_{\rm ex}^{(1)}&=\int\frac{\mathrm{d}^2q}{(2\pi)^2}p_{\bm q}
=\frac{1}{4\pi^2v^2\tau_Q}.
\end{align}
Including both valleys,
\begin{equation}
\boxed{n_{\rm ex}=\frac{N_D}{4\pi^2v^2\tau_Q},\qquad N_D=2.}
\end{equation}
A spatially moving domain wall avoids a simultaneous gap closing throughout the sample: the mass changes sign only near the moving interface, while the remote bulk remains gapped.

\begin{figure}[t]
\centering
\includegraphics[width=0.82\linewidth]{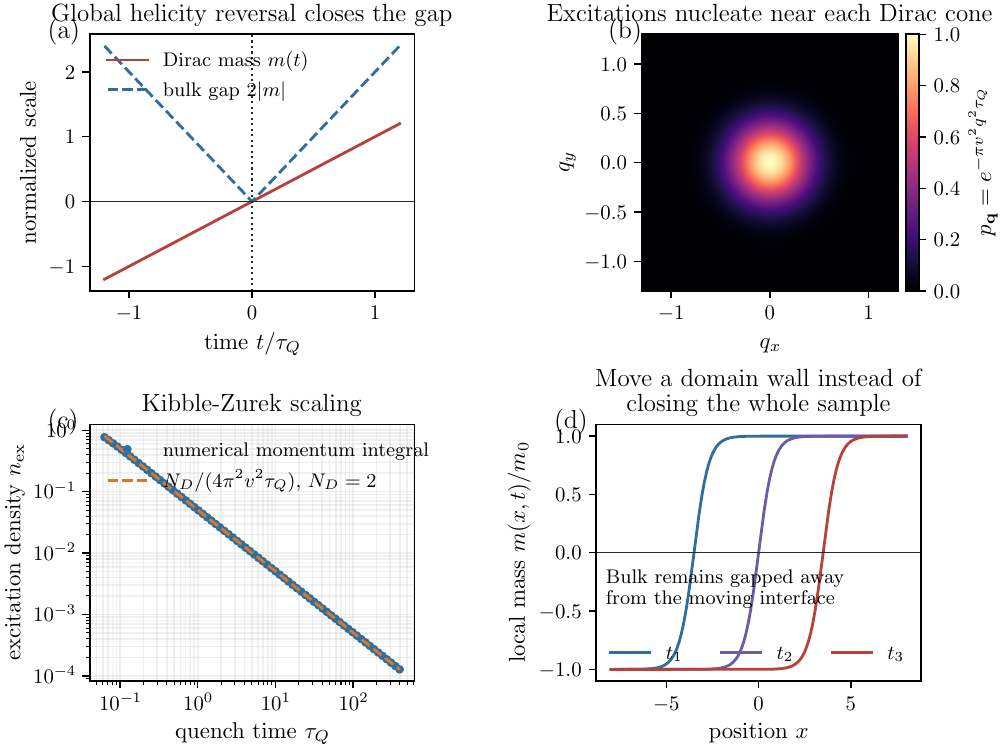}
\caption{\label{fig:switching}
Helicity switching is a dynamical topological transition. Spatial rewriting localizes the unavoidable gap closure.
 (a) A uniform reversal changes the sign of the Dirac mass and necessarily closes the bulk gap. (b) Excitations are generated near each Dirac cone with the Landau--Zener probability shown. (c) The integrated two-cone excitation density follows $n_{\rm ex}\propto\tau_Q^{-1}$. (d) Translating a domain wall localizes the gap closure at a moving interface while leaving the distant bulk gapped.}
\end{figure}

Figure~\ref{fig:switching} contrasts the two protocols. A uniform reversal must pass through zero mass everywhere. Slower ramps only narrow the excited region around each valley, producing the $\tau_Q^{-1}$ law. Translating a preexisting wall instead confines the one-dimensional gapless region to the moving interface and trades global defect production for smooth transport relative to the interface localization length.

\CGBStandaloneEnd

\let\CGBSupplementBodyOnly\undefined
\let\CGBStandaloneEnd\undefined
\let\CGBFigureFour\undefined
\let\CGBFigureFive\undefined

\end{widetext}

\end{document}